\documentclass[twocolumn,showpacs,preprintnumbers,amsmath,amssymb,epsfig,widetext]{revtex4-1}

\usepackage{graphicx}
\usepackage{dcolumn}
\usepackage{bm}
\usepackage{epsfig}
\usepackage{color}

\def\fun#1#2{\lower3.6pt\vbox{\baselineskip0pt\lineskip.9pt
  \ialign{$\mathsurround=0pt#1\hfil##\hfil$\crcr#2\crcr\sim\crcr}}}
\def\simgt{\mathrel{\lower0.6ex\hbox{$\buildrel {\textstyle >}
 \over {\scriptstyle \sim}$}}}
\def\simlt{\mathrel{\lower0.6ex\hbox{$\buildrel {\textstyle <}
 \over {\scriptstyle \sim}$}}}

\input epsf

\newcommand{\hompc}{\,h\,{\rm Mpc}^{-1}}
\newcommand{\mpcoh}{\,h^{-1}\,{\rm Mpc}}
\newcommand{\ompc}{\,{\rm Mpc}^{-1}}
\newcommand{\mpc}{\,{\rm Mpc}}
\newcommand{\mnras}{MNRAS}

\newcommand{\dec}{\rm dec}

\def\be{\begin{equation}}
\def\ee{\end{equation}}
\def\ba{\begin{eqnarray}}
\def\ea{\end{eqnarray}}

\def\nn{\nonumber}

\begin{document}

\preprint{}

\title{Coherent Combination of Baryon Acoustic Oscillation Statistics and Peculiar Velocity Measurements from Redshift Survey}
 
\author{Yong-Seon Song}
\email{ysong@kias.re.kr}
\affiliation{Korea Institute for Advanced Study, Dongdaemun-gu, Seoul 130-722, Korea}

\date{\today}

\begin{abstract}
New statistical method is proposed to coherently combine Baryon Acoustic Oscillation statistics (BAO) and peculiar velocity measurements exploiting decomposed density--density and velocity--velocity spectra in real space from the observed redshift distortions in redshift space, 1) to achieve stronger dark energy constraints, $\sigma(w)=0.06$ and $\sigma(w_a)=0.20$, which are enhanced from BAO or velocity measurements alone, and 2) to cross--check consistency of dark energy constraints from two different approaches; BAO as geometrical measurements and peculiar velocity as large scale structure formation observables. In addition to those advantages, as power spectra decomposition procedure is free from uncertainty of galaxy bias, this simultaneous fitting is an optimal method to extract cosmological parameters without any pre--assumption about galaxy bias.
\end{abstract}

\pacs{draft}


\maketitle

\section{Introduction}

The evolution of large scale structure, as revealed in the clustering of galaxies observed in wide--deep redshift surveys, has been one of the key cosmological probes. Structure formation is driven to grow by a competition between gravitational attraction and the expansion of space-time. This enables us to test our model of gravity at cosmological scales as well as the expansion history of the Universe. Although galaxies are not an unbiased tracer of the large scale matter distribution predictable from linear gravitational theory, various statistical methods using galaxy redshift survey have been developed to probe the nature of cosmic acceleration such as dark energy or modified gravity. Acoustic peak structure imprinted on density field inhomogeniety provides mapping to the large scale power spectra in transverse and longitudinal directions. Before decoupling, the photons and baryons are tightly coupled to form plasma, and overdensity with baryon acoustic oscillation feature (BAO) develops through the competition between outward pressure support of radiation and gravitational attraction. After decoupling, as baryonic matter is influenced only by gravitational force at large scale, the baryons and dark matter instabilities form a configuration including predictable BAO pattern extending from the sound horizon measured by CMB at decoupling epoch. This appears to be a standard ruler tracing the relation between redshift and expansion rate~\cite{Seo:2003pu,Blake:2003rh,Linder:2003ec,Hu:2003ti,2004MNRAS.348..250C,Matsubara:2004fr,Amendola:2004be,Blake:2004tr,Glazebrook:2005mb,Dolney:2004va,Seo:2007ns,Jeong:2008rj,Simpson:2009zj,Samushia:2010ki}. Additionally, as galaxies are expected to act nearly as test particles within the cosmological matter flow, the motions of galaxies carry an imprint on the rate of growth of large-scale structure and allows us to constrain cosmological models~\cite{Wang:2007ht,Song:2008qt,McDonald:2008sh,Bean:2010zq,Jennings:2010uv,Taruya:2010mx,Song:2010kq}. Despite the intensive studies of both methods constraining cosmological parameters, the coherent way to combine both has not yet been fully optimized. 

Both different statistics can be combined coherently exploiting decomposed density--density and velocity--velocity spectra in real space from the observed redshift distortions in redshift space. The Fisher matrix exposition was developed~\cite{White:2008jy}, assuming strict functional forms for the power spectra or allowing them to float freely. As expected the constraints are tightest when theoretical investigations can provide good priors for the form of distorted power spectra in redshift space, but even relatively conservative assumptions suggest that percent level of decomposition should be possible with future surveys~\cite{White:2008jy,Song:2010bk}. The cosmological distance errors are achievable using BAO preserved on decomposed density--density spectra, which is degraded a few factors of order from BAO of full observed power spectra. Decrement is compensated by adding cosmological constraints from the decomposed velocity--velocity spectra utilizing the growth of large-scale structure as imprinted on dynamics of galaxies observed in large redshift surveys. Tighter constraints are achieved by this combination as well as dark energy constraints can be cross--checked by two different alternative approaches; distance measures and large scale structure formation observables. 

The detailed formalism is presented in the next section. The Fisher matrix analysis to decompose spectra is briefly reviewed, then BAO statistics is redefined with the decomposed density--density spectra. Coherent formalism to combine BAO and decomposed velocity--velocity spectra is introduced, and advantages are discussed to gain stronger constraints on dark energy models with future surveys. For illustration, a fiducial $\Lambda$CDM cosmology is assumed with $\Omega_{\rm m}=0.24$, $h=0.72$, $n=0.96$ and $A_S^2=2.41\times 10^{-9}$ (in good agreement with a variety of observations) when computing specific predictions for future surveys.

\section{Coherent approach to combine BAO and peculiar velocity}

The formalism is briefly reviewed to forecast errors on galaxy--galalxy density spectra, $P_{gg}$, and velocity--velocity spectra, $P_{\Theta\Theta}$ ($\Theta$ is divergence of velocity map in unit of $aH$, $\Theta=\vec{\nabla}\cdot\vec{v}/aH$) from the observed spectra, $\tilde{P}$, of redshift surveys. We explain in detail how to estimate error forecast of BAO from the decomposed density--density spectra as well as how to combine decomposed peculiar velocity measurement coherently.

\subsection{Decomposition of the observed power spectra in redshift space}

The observed two-point correlation function in redshift space is decomposed into spectra of density fluctuations and peculiar velocity fields in real space~\cite{White:2008jy}. The power spectra in redshift space, $\tilde{P}$, are given by,
\begin{eqnarray}
\tilde{P}(k,\mu,z) &=& \big\{P_{gg}(k,z)
   + 2\mu^2r(k)\left[P_{gg}(k,z)P_{\Theta\Theta}(k,z)\right]^{1/2}\nonumber\\
  &+& \mu^4P_{\Theta\Theta}(k,z)\big\}G_{\rm FoG}(k,\mu,\sigma_z) 
\end{eqnarray}
where $\mu$ denotes the cosine of angle between orientation of two point correlation and the line of sight. The cross-correlation coefficient $r(k)$ is defined as $r(k)\equiv P_{g\Theta}/\sqrt{P_{gg}P_{\Theta\Theta}}$. The density and velocity fields are highly correlated for $k<0.1\,h\,{\rm Mpc}^{-1}$ thus we assume  that the density and velocities are perfectly correlated, $r(k)\sim 1$. The density-velocity cross-spectrum becomes the geometric mean of the two auto-spectra to leave only two free functions, $P_{gg}$ and $P_{\Theta\Theta}$. Uncertainty in the observed redshifts is modeled by a line-of-smearing of the structure using fitting function $G_{\rm FoG}=e^{-k^2\sigma_z^2\mu^2}$ where $\sigma_z$ denotes one--dimensional velocity dispersion (FoG: Finger-of-God effect).

The accuracy of decomposition of $P_{gg}$ and $P_{\Theta\Theta}$ out of $\tilde{P}$ is estimated using Fisher matrix analysis determining the sensitivity of a particular measurement. Fisher matrix for this decomposition, $F_{\alpha\beta}^{\rm dec}$, is written as,
\ba\label{eq:Fdec}
F_{\alpha\beta}^{\dec}=\int\frac{\partial\tilde{P}(\vec{k})}{\partial p_{\alpha}}\frac{V_{\rm eff}(\tilde{P})}{\tilde{P}(\vec{k})^2}\frac{\partial\tilde{P}(\vec{k})}{\partial p_{\beta}}\frac{d^3k}{2(2\pi)^3}w_{\rm FoG}(\vec{k})\,,
\ea
where $p_{\alpha}=(P_{gg}^{\dec},P_{\Theta\Theta}^{\dec})$. The effective volume $V_{\rm eff}(\tilde{P})$ is given by,
\ba
V_{\rm eff}(\tilde{P})=\left[\frac{n\tilde{P}}{n\tilde{P}+1}\right]^2V_{\rm survey}\,,
\ea
where $n$ denotes galaxy number density, here $n=5\times 10^{-3}(\mpcoh)^{-3}$. Comoving volume, $V_{\rm survey}$, given by each redshift shell from $z=0$ to 2 with $\Delta z=0.2$ ($f_{sky}=1/2$) is written as, 
\ba
V_{\rm survey}=f_{sky}\frac{4\pi}{3}(D_{\rm outer}^3-D_{\rm inner}^3)\,,
\ea
where $D_{\rm outer}$ and $D_{\rm inner}$ denote comoving distances of outer and inner shell of the given redshift bin respectively. The weight function $w_{\rm FoG}(k,\mu)$ is given by
\begin{equation}
  w_{\rm FoG}(k,\mu)=
  \exp{\left[-\frac{(G_{\rm FoG}-1)^2}{\sigma_{\rm th}^2}\right]}\,,
\end{equation}
where $G_{\rm FoG}$ is the finger-of-god suppression factor and $\sigma_{\rm th}$ is a threshold value indicating our confidence in the accuracy of the FoG model. We fix $\sigma_{\rm th}=0.1$ indicating trust of clearance of non-linear effect up to 10$\%$ contamination of the factor beyond Kaiser limit. Derivative terms in Eq.~(\ref{eq:Fdec}) are given by,
\begin{eqnarray}
  \frac{\partial \ln \tilde{P}(k_i,\mu,z_j)}{\partial P_{gg}^{\dec}(k_i,z_j)}
  &=& \frac{1}{\tilde{P}(k_i,\mu,z_j)}
  \left[1 + \mu^2
  \sqrt{\frac{P_{\Theta\Theta}(k_i,z_j)}{P_{gg}(k_i,z_j)}} 
  \right] \nonumber \\
  \frac{\partial\ln \tilde{P}(k_i,\mu,z_j)}{\partial P_{\Theta\Theta}^{\dec}(k_i,z_j)}
  &=&\frac{\mu^2}{\tilde{P}(k_i,\mu,z_j)}
  \left[\sqrt{\frac{P_{gg}(k_i,z_j)}{P_{\Theta\Theta}(k_i,z_j)}}+\mu^2\right]\,.
\end{eqnarray}

\begin{figure}[t]
 \begin{center}
 \epsfysize=3.truein
   \epsffile{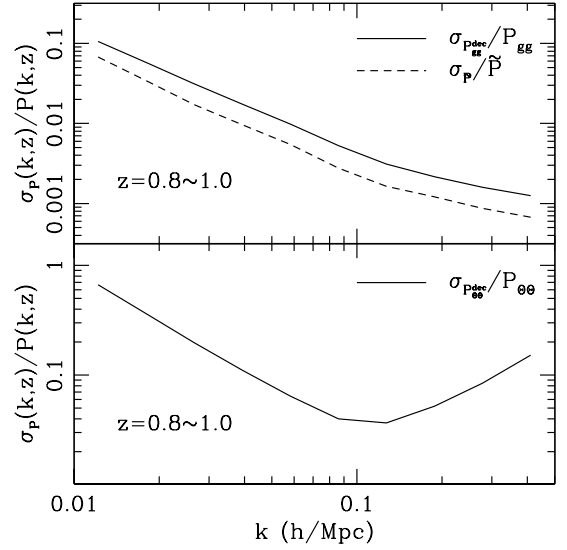}
   \caption{\footnotesize Fractional errors of power spectra are presented in both panels. {\it Upper panel:} solid and dash curves represent fractional errors of decomposed $P_{gg}^{\rm dec}$ and observed $\tilde{P}$. {\it Lower panel:} solid curve represents fractional errors of decomposed $P_{\Theta\Theta}$. Here $\sigma(P)$ denotes the error on power spectra $P$, and $\sigma(P)/P$ denotes the fractional error of it.} 
\label{fig:sigma}
\end{center}
\end{figure}

\subsection{Variances of power spectra}

The detectability of galaxy--galaxy density spectra in redshift space, $\tilde{P}$, is given by,
\ba\label{eq:sigtP}
\frac{1}{\sigma_{\tilde{P}}^2(k_i,z_j)}&=&\int^{1}_{-1}\int^{k^i_{\rm max}}_{k^i_{\rm min}}\frac{V_{\rm survey}(z_j)}{[\tilde{P}(k,\mu,z_j)+1/n]^2}\frac{k^2dkd\mu}{8\pi^2}\,,
\ea
where $'j'$ denotes redshift bin from $z=0$ to 2, and $'i'$ denotes $k$ space bin running from $k=0.01\hompc$ to $0.5\hompc$, i.e. integration of $k$ in each $k_i$ bin from lower to upper bounds denoted by $k^i_{\rm min}$ and $k^i_{\rm max}$ respectively.


The detectability of galaxy--galaxy density spectra in real space, $P_{gg}$, can be estimated from decomposition Fisher matrix $F_{\alpha\beta}^{\rm dec}$. The $gg$--component of the inverse matrix of $F_{\alpha\beta}^{\rm dec}$ gives the variance of $P_{gg}^{\dec}$ as,
\ba\label{eq:sigPgg}
\sigma_{P_{gg}^{\dec}}(k_i,z_j) = \sqrt{\left(F^{\rm dec\,-1}\right)_{gg}(k_i,z_j)}\,.
\ea
In the upper panel of Fig.~\ref{fig:sigma}, solid curve represents fractional variance of $P_{gg}^{\dec}$ at the same redshift bin $0.8<z<1.0$. The decomposition procedure to transfrom measured galaxy--galaxy density spectra in redshift space into real space degrades detectability, but is not very significant about a factor of 1.8.

Variance of decomposed velocity--velocity spectra is derived from $\Theta\Theta$--component of the inverse matrix of $F_{\alpha\beta}^{\rm dec}$,
\ba\label{eq:sigPtt}
\sigma_{P_{\Theta\Theta}^{\dec}}(k_i,z_j) = \sqrt{\left(F^{\rm dec\,-1}\right)_{\Theta\Theta}(k_i,z_j)}\,.
\ea
In the lower panel of Fig.~\ref{fig:sigma}, solid curve represents fractional variance of $P_{\Theta\Theta}^{\dec}$ at redshift bin of $0.8<z<1.0$. As $\sigma_{\rm th}$ is set to be 0.1, the detectability is significantly weakened at $k>0.1 \hompc$. It is observed V--shape of variance pointed around threshold scale of $k=0.1\hompc$. Due to the short range of detectability, probing BAO structure using $P_{\Theta\Theta}$ proves to be challenge, but, as $P_{\Theta\Theta}$ is not a biased tracer, the measured amplitude can be used to constrain cosmological parameters.

Threshold limit representing cutting--off scale of $k$ due to the unpredictable non--linear physics, $\sigma_{\rm th}$,  reduces detectability of spectra shown in Fig.~\ref{fig:sigma}, but differently affects on $P_{gg}^{\dec}$ and $P_{\Theta\Theta}^{\dec}$. The introduced $\sigma_{\rm th}$ calibrates Fisher matrix integrand at the combination of $k\mu$ not $k$. The decomposition of $P_{gg}$ is less affected, as it is most determined at $\mu\rightarrow 0$ limit, while $P_{\Theta\Theta}$ is better decomposed at $\mu\rightarrow 1$ limit. Poorer detectability of $P_{\Theta\Theta}$ at $k>0.1\hompc$ is due to increasing FoG calibration at $\mu\rightarrow 1$ limit.

\subsection{Formalism of BAO Fisher matrix}

Cosmological constraints are forecasted from a measured position of BAO in the three dimensional power spectra following the formalism of~\cite{Seo:2007ns}. The baryonic part of the power spectrum can be modeled as 
\begin{equation}
  P_{\rm b}(k) \sim \frac{\sin{ks_0}}{ks_0}\exp{\left[-\left(\frac{k}{k_{\rm
  Silk}}\right)^{1.4}\right]},
  \label{eq:pb}
\end{equation}
where $s_0$ is the sound horizon at drag epoch and $k_{\rm Silk}$ is the Silk damping scale. The Silk damping scale can be accurately fitted by $k_{\rm Silk} = 1.6(\Omega_{\rm b}h^2)^{0.52}(\Omega_{\rm m}h^2)^{0.73}[1+(10.4\Omega_{\rm m}h^2)^{-0.95}]h^{-1}\hompc$. 

As in~\cite{Seo:2007ns}, Eq.~(\ref{eq:pb}) is multiplied by additional Gaussian functions to account for the erasure of information due to nonlinear evolution. The final $P_{\rm b}$ is given by
\begin{eqnarray}
  P_{\rm b}(k) =
  \sqrt{8\pi^2}A_0P_{0.2}\frac{\sin{ks_{0}}}{ks_{0}} G_{BAO}^{1/2}(k)\,,
  \label{eq:pb2}
\end{eqnarray}
where $P_{0.2}$ is galaxy power spectrum at $k = 0.2 \hompc$, $A_0$ is a normalization factor, $\Sigma_{\rm nl}$ models the loss of information due to nonlinear growth, and $G_{BAO}(k)$ is defined as,
\ba\label{eq:GBAO}
G_{BAO}(k)\equiv e^{-2(k/k_{\rm Silk})^{1.4}-k^2\Sigma_{\rm nl}^2}\,.
\ea
Eq.~(\ref{eq:pb}) is compared with the fitting formula of Seo and Eisenstein to estimate $A_0$. The fiducial cosmology, as chosen in this research, has an analytical estimate of $A_0 = 0.46$. As in~\cite{Seo:2007ns} the numerical values $\Sigma_{\rm nl}= \Sigma_0g_{\delta_m}$, where $g_{\delta_m}$ is the growth function normalized as $g_{\delta_m}(a)=a$ at early time, and $\Sigma_0 = 11.0\hompc$ for the cosmology with $\sigma_8 =0.8$. When the physical value of the sound horizon is known to high precision from CMB measurements the error on $s_{\rm 0}$ is equivalent to the error on $D_V\equiv(D_A^2/H)^{1/3}$.

\begin{figure}[t]
 \begin{center}
 \epsfysize=3.truein
   \epsffile{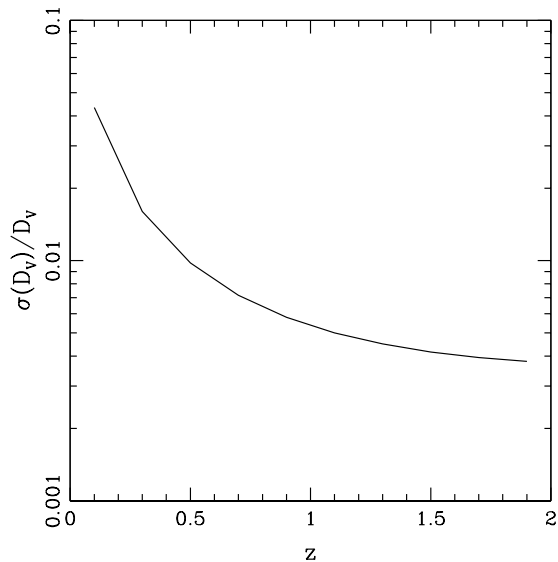}
   \caption{\footnotesize Fractional errors on $D_V(z)$ from $P_{gg}^{\dec}$ BAO are represnedted by solid curve.}
\label{fig:DV}
\end{center}
\end{figure}

Centroid approximation is available for $P_{gg}^{\dec}$, as it is decomposed into spherically symmetric Fourier space. BAO Fisher matrix from $P_{gg}^{\dec}$ is given by,
\begin{eqnarray}\label{eq:disFgg}
 F_V^{P_{gg}^{\dec}}(z_j) \sim 4\pi^2A_0^2 \displaystyle\sum_{i=1}^{N_k} \frac{P_{0.2}^2}{\sigma_{P_{gg}^{\dec}}(k_i,z_j)^2}G_{BAO}(k_i)\,,
\end{eqnarray}
where $N_k$ is the total number of $k$ bin upto $k=0.1\hompc$. The variance of $D_V$ is simply given by $\sigma(D_V^P)=1/\sqrt{F_V^{P}}$ of $P_{gg}^{\dec}$. In Fig.~\ref{fig:DV}, fractional errors of $D_V$ calculated from $P_{gg}^{\dec}$ only are presented. 

\subsection{Tight constraints on dark energy from $P_{\Theta\Theta}^{\dec}$}

The observed power spectra depend not only on dark energy parameters but the entire matter content and primordial power spectrum. We do not assume these quantities to be known. Instead, it is assumed that CMB data are available to constrain them.  The CMB power spectra included in our analyses are the $C_l$ spectra of temperature--temperature, temperature--polarization and polarization--polarization. Cosmological parameter space is given by ($w$, $w_a$, $\omega_b$, $\omega_m$, $\theta_S$, $A_S$, $n_S$, $z_{reion}$). 

With CMB temeprature fixed at $T_{\rm CMB}=2.726\,{\rm K}$, the sound horizon at last scattering surface, $D_S(z_{lss})$, is given in $\mpc$ unit by $\omega_m$ and $\omega_b$ measured by CMB acoustic peak structure. With the measured angular size of sound horizon at last scattering surface, the angular diameter distance to the last scattering surface is also given in $\mpc$ unit. The geometrical factor determination in $\mpc$ unit is converted into the Fourier space dimension of $k$ in $\ompc$ unit. As primordial spectra are tightly constrained by CMB physics, the pivoting scale of primordial spectra $k_{p}$ is given in $\ompc$ unit. In addition to this, as the shape of spectra in terms of $k$ (transfer function $T_{\Phi}$) is tightly determined by parameters measured by CMB physics, we express $T_{\Phi}$ in terms of $k$ in $\ompc$ unit.

In redshift survey, measured redshift of objects is converted using distance in $\mpcoh$ unit in which mapping is less affected by most uncertain distance parameter of the Hubble constant. The velocity--velocity power spectra are given by in terms of $k$ in $\hompc$ unit~\cite{Song:2010kq},
\ba
P_{\Theta\Theta}(k,a)=\frac{8\pi^2}{25}\frac{k}{H_*^4\Omega_m^2}A_S^2\left(\frac{kh}{k_{p}}\right)^{n_S-1}T_{\Phi}^2(kh)g^2_{\Theta}(a)\nn\,.
\ea
Variation of dark energy parameter impact on $P_{\Theta\Theta}(k,a)$ becomes complicated when we consider the combined constraints from CMB and redshift survey in which $k$ dependent quantities are measured in different units. First, multiplication factor becomes $k/H_*^4\Omega_m^2$ which changes due to floating of $\Omega_m$ and $h$. Second, primordial tilt factor changes as $k$ and $k_{p}$ are determined by different units. Third, transfer function factor $T_{\Phi}$ is unmodified in terms of $kh$ with varying dark energy parameters, as decay due to silk damping is well determined by cosmological parameters measured at last scattering surface with CMB physics. But it is altered due to using unit of $\mpcoh$ instead of $\rm Mpc$ for distance measure of redshift survey. Finally, growth factor changes.

\begin{figure}[t]
 \begin{center}
 \epsfysize=3.truein
   \epsffile{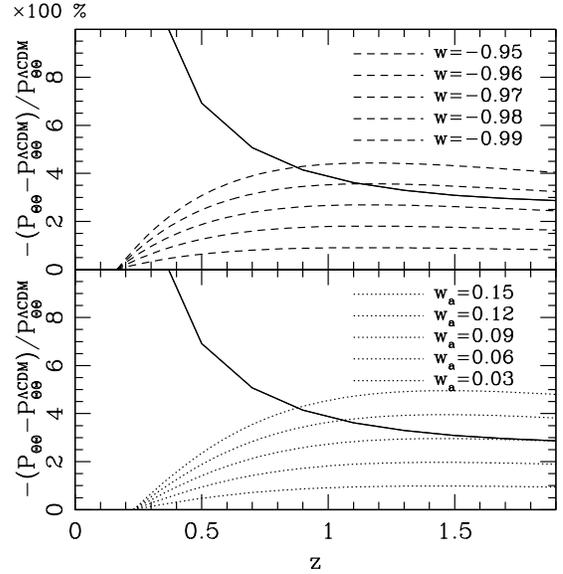}
   \caption{\footnotesize Solid curves represent fractional error of $P_{\Theta\Theta}^{\dec}$ at $k$ range of $0.07<k<0.1\hompc$ in both panels. {\it Upper panel:} dash curves represent percentage variation of $P_{\Theta\Theta}^{\dec}$ with different $w$ from $w=-1$, $w=-0.95,-0.96,-0.97,-0.98,-0.99$ from top to bottom. {\it Lower panel:} dotted curves represent percentage variation of $P_{\Theta\Theta}^{\dec}$ with different $w_a$ from $w_a=0$, $w=0.15,0.12,0.09,0.06,0.03$ from top to bottom. Here, $k=0.86\hompc$ for both panels.}
\label{fig:dpk}
\end{center}
\end{figure}

In Fig.~\ref{fig:dpk}, total change of $P_{\Theta\Theta}(k=0.86\hompc,a)$ is presented with varying $w$ (upper panel) and $w_a$ (lower panel). If degeneracy is not considered between $w$ and $w_a$, tight constraints on dark energy paramters are observed. Fourier space is split into 10 bins from $k=0.01\hompc$ to $0.5\hompc$. This figure is a representation by one of 10 $k$ bins chosen at $0.07<k<0.1\hompc$. 

\subsection{Coherent combination of $P_{gg}^{\dec}$ and $P_{\Theta\Theta}^{\dec}$}

To calculate the expected parameter errors we make a first order Taylor expansion of the parameter dependence including CMB, BAO and decomposed $P_{\Theta\Theta}$. The expected parameter errors are easily calculated from the inversed covariance matrix of the Fisher matrix using this `linear response' approximation. The linear response approximation is improved and susceptibility to numerical error is reduced with a careful choice of the parameters including $\theta_S$. $F^{\rm CMB}_{mn}$ denotes a Fisher matrix of CMB physics equivalent to combination of distance measure at last scattering surface and primordial parameter constraints, thus exploiting unlensed CMB power spectra only. We use Planck experiment specs at three frequencies of 100, 143 and 217 GeV. Each channel has the resolution of 9.1, 7.2, 5.0 arcmin , the temperature sensitivity of 5.5, 6.0 and 13 $\mu$K and the polarization sensitivity of $\infty$, 11, 27 $\mu$K respectively. We calculate $F^{\rm CMB}_{mn}$ using the method developed in our previous paper~\cite{Kaplinghat}.

\begin{figure}[t]
 \begin{center}
 \epsfysize=3.truein
   \epsffile{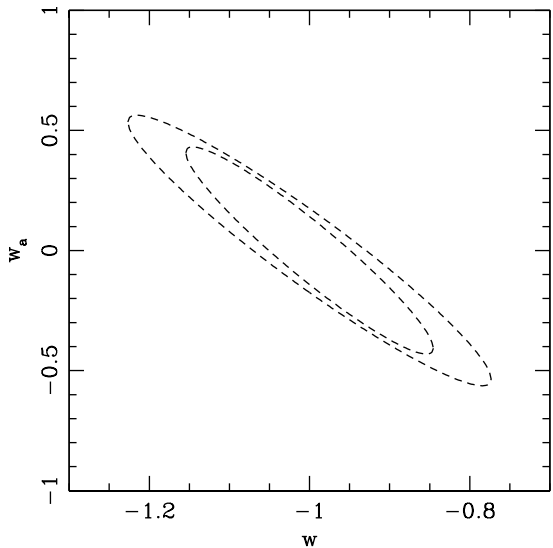}
   \caption{\footnotesize We present the fractional error dependence of dark energy constraints from BAO in the upper panel of Fig.~\ref{fig:sigma}; outer dash contour represents dark energy constraint from the fractional error of $P_{gg}^{\rm dec}$ (solid curve in the upper panel of Fig.~\ref{fig:sigma}), and inner dash contour represents dark energy constraint from the fractional error of $\tilde{P}_{gg}$ (dash curve in the upper panel of Fig.~\ref{fig:sigma}).}
\label{fig:con_BAO}
\end{center}
\end{figure}

\begin{figure}[t]
 \begin{center}
 \epsfysize=3.truein
   \epsffile{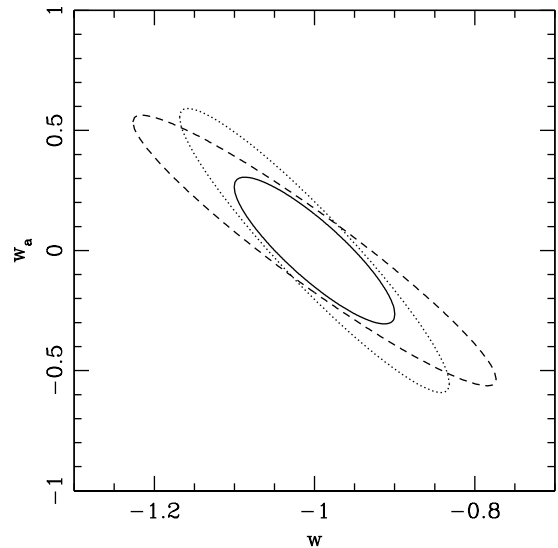}
   \caption{\footnotesize Constraints on dark energy parameters are presented; dark energy constraint from $P_{gg}^{\rm dec}$ BAO (dash contour), $P_{\Theta\Theta}$ (dotted contour) and coherent combination of both (solid contour).}
\label{fig:con}
\end{center}
\end{figure}

Constraint on distance measure $D_V$ by BAO is converted into cosmological parameter constraints using the following Fisher matrix formalism,
\ba
F^{\rm D_V^{P_{gg}^{\dec}}}_{mn}=\sum_{z_j=0}^2\frac{\partial D_V^{P_{gg}^{\dec}}(z_j)}{\partial x_m}\frac{1}{\sigma(D_V^{P_{gg}^{\dec}})^2}\frac{\partial D_V^{P_{gg}^{\dec}}(z_j)}{\partial x_n}\,,
\ea
where indices $m$ and $n$ denote elements in cosmological parameter space. Dark energy constraints from $F^{\rm D_V^{P_{gg}^{\dec}}}_{mn}+F^{\rm CMB}_{mn}$ are shown to be the outer dash contour in Fig.~\ref{fig:con_BAO} and the dash contour in Fig.~\ref{fig:con}, $\sigma(w)=0.15$ and $\sigma(w_a)=0.37$, which is weaker than dark energy constraint from BAO of full observed spectra, $\sigma(w)=0.10$ and $\sigma(w_a)=0.30$ (the inner dash contour in Fig.~\ref{fig:con_BAO}). Here error estimation $\sigma$ denotes 68$\%$ statistical significance. 

While $D_V^{P_{gg}^{\dec}}$ constrains dark energy by distance measure, dark energy is probed by signature on the large scale structure growth of peculiar velocity measurements $P_{\Theta\Theta}^{\dec}$. Shown in the bottom panel of Fig.~\ref{fig:sigma} for fractional errors of $P_{\Theta\Theta}$, the effective range of $k$ is too narrow to observe the baryonic acoustic peak structure.  However unlike $P_{gg}^{\dec}$, it is an unbiased quantity that is predictable with a given set of cosmological parameters. Fisher matrix is given by,
\ba\label{eq:fishtt}
F^{\Theta\Theta}_{mn}=\sum_{k=1}^{N_k^{\rm cut}}\sum_{z_j=0}^2\frac{\partial P_{\Theta\Theta}(k_i,z_j)}{\partial x_m}\frac{1}{\sigma_{P_{\Theta\Theta}^{\dec}}(k_i,z_j)^2}\frac{\partial P_{\Theta\Theta}(k_i,z_j)}{\partial x_n}\nn\,.
\ea
At scales greater than $k=0.1\hompc$, the uncertainty increases due to the unknown non--linear physics causing the FoG effect and the violation of assumption of perfect cross--correlation between density--density and velocity--velocity spectra~\cite{Song:2010bk}. It was shown that the decomposed $P_{\Theta\Theta}$ starts to be biased as scale approaches to $k=0.1\hompc$. Cut--off scale needs to be introduced at $k_{\rm cut}=0.1\hompc$ in order to skim away contributions contaminated from non--linear physics at $k>0.1\hompc$.

In Fig.~\ref{fig:con}, dotted contour represents dark energy constraint from $P_{\Theta\Theta}^{\dec}$. Tight contraints achieved using $P_{\Theta\Theta}^{\dec}$ is worth being investigated. 
At last redshift bin of $1.8<z<2.0$, dark energy constraints can be read from Fig.~\ref{fig:dpk} about $\sigma(w)\sim 0.035$ or $\sigma(w_a)\sim 0.09$ which confirms the results from Eq.~\ref{eq:fishtt}. If fitting is extended to all redshift bins then $\sigma(w)=0.015$ or $\sigma(w_a)= 0.04$ (derived from Eq.~\ref{eq:fishtt}). This still agrees with that is observed in Fig.~\ref{fig:dpk}. But simultaneous determination of $w$ and $w_a$ becomes less tight as $\sigma(w)=0.18$ and $\sigma(w_a)= 0.50$, because of degenerate $P_{\Theta\Theta}$ variation in terms of $w$ and $w_a$. The results shown in Fig.~\ref{fig:con} are given after full marginalization with all other cosmological parameters.

Then coherent combination of BAO and peculiar velocity measurements can be simply given by
\ba
F^{\rm total}_{mn}=F^{\rm D_V^{P_{gg}^{\rm dec}}}_{mn}+F^{\Theta\Theta}_{mn}+F^{\rm CMB}_{mn}\,.
\ea
Solid contour of Fig.~\ref{fig:con} represents dark energy constraints from $F^{\rm total}_{mn}$, $\sigma(w)=0.06$ and $\sigma(w_a)= 0.20$, which is tighter than any other case considered in this paper.

\section{Conclusion}

The known galaxy bias has always been considered an obstacle utilizing redshift surveys for cosmological purpose. If just peak structure is considered as a standard ruler at the given targeted redshift, cosmological information is extractable with less dependence on bias uncertainty. However enriched information of large scale structure is not fully exploited. Previously, decomposed contributions of density--density and velocity--velocity spectra were studied in the observed redshift distortion using anisotropic feature aligned along the line of sight. In this paper, it is proposed to coherently combine both decomposed spectra for constraining cosmological parameters (still in bias uncertainty independent way) to enhance dark energy constraints. 

What is implemented in this new method is to cross--check constraints on dark energy in two different approaches; distance measure by BAO and large scale stucture growth factors by peculiar velocity measurements. There are many different theoretical models proposed to generate cosmic acceleration. It is essential to test those theories in many different observational methods. When we decompose spectra into density--density and velocity--velocity pairs, we can test theoretical models using geometrical tools as well as using large scale structure formation. In addition, dark energy constraints are enhanced from performing BAO alone using density--density spectra before decomposition. In this sense, our approach is optimizing way to exploit redshift surveys for the purpose of constraining dark energy.

It is important to test consistency between geometrical measures and structural observable. As general relativity can be modified at large scales. One of vital mission of dark energy probe is to distinguish two different class of models; modifying energy momentum components or modifying gravity. Coherent motion measurements provide us with a way of tracing Newtonian force sourcing dynamics of gravitational instability. Our new statistical methods assist us testing modified gravity type models too.

It has not yet been fully tested whether the assumption of preservation of baryon acoustic peak structure is true or not, although baryon acoustic structure is not expected to be lost in the decomposition procedure. It is necessary to test this assumption using N-body simulation. For decomposition of velocity measurements, it was tested whether there was velocity bias or not. However at the same time, it was known that velocity--velocity spectra are most sensitive to be extracted without systematic uncertainty. In the future, higher resolution experiments will be launched, and it would be interesting whether velocity bias is not induced yet. 

\section*{Acknowledgments}

We would like to thank Lado Samuthia for discussion about BAO Fisher matrix, Eric Linder for helpful comments in details, and the referee for improving the manuscript. We thank Korea Institute for Advanced Study for providing computing resources (KIAS linux cluster system) for this work.


\end{document}